\documentclass[twocolumn,showpacs,amsmath,amssymb]{revtex4}
\usepackage{graphicx}
\usepackage{color} 

\begin{document}

\title{Spin-glass-like behavior caused by Mn-rich Mn(Ga)As nanoclusters in GaAs}
\author{C. H. Chang and T. M. Hong}
\affiliation{Department of Physics, National Tsing Hua University, Hsinchu 30043, Taiwan, Republic of China}
\date{\today}

\begin{abstract}
We simulate the indirect exchange interaction between  Mn-rich Mn(Ga)As nanoclusters in GaAs by analytical means. In contrast to the conventional Ruderman-Kittel-Kasuya-Yosida (RKKY) formula which considers the mediation by the carriers in the medium, we also include the contribution from those inside the clusters. Since the carrier concentration is higher in the clusters, this modification allows the RKKY oscillation to change sign. Consequently, while the previous approach only favors ferromagnetism for this system, an antiferromagnetic coupling is in fact possible. Since the Mn-rich Mn(Ga)As nanoclusters are naturally formed and bound to have different sizes, their spin orientation is likely to be frustrated due to mixed preferences from different neighbors. We argue that this is likely the source of the spin-glass-like behavior which plagues this system. By tuning the size and narrowing its distribution, normal ferromagnetism can be restored with an Curie temperature higher than previously thought.
\end{abstract}


\maketitle
%%%%%%%%%%%%%%%%%%%%%%%%%%%%%%%%%%%%%%%%%%%%%%%%%%%%%%%%%%%%%%%%%%%%%%%%%%%%%%%%
%%%%%%%%%%%%%%%%%%%%%%%%%%%%%%%%%%%%%%%%%%%%%%%%%%%%%%%%%%%%%%%%%%%%%%%%%%%%%%%%

DMS (Diluted Magnetic Semiconductors) have been hailed as a potential spintronic device to integrate
the computing power of CPU and the storage ability of the hard disc\cite{alan,dietl,zutic}.
By intimately coupling them, we hope to come up with a device that can compute while recording, and vice versa.
Besides the ion-implanted samples, nanoclusters natually formed or embedded in a semiconducting matrix have attracted increasing interest due to
their potential of raising the Curie temperature and rich magnetic properties\cite{cl1,cl2,cl3}, e.g., an enhanced magneto-optical and magnetotransport response that can be applied to the magneto-optical spatial light modulators for volumetric recording.
When nanoparticles are small enough to enter the superparamagnetic regime, they
behave like a giant spin without a permanent magnetic moment.

The indirect coupling between these magnetic nanoparticles has been studied in the context of GMR (Giant Magnetoresistance)\cite{toro} for a metallic medium. Note that, although the underlying mechanism is expected to follow the theory by Ruderman, Kittel, Kasuya, and Yosida (RKKY), how the actual coupling between nanoparticles differs from the conventional one, namely, between point impurities is not clear. Previous workers assumed the validity of conventional RKKY formula and concentrated on the mathematical difficulty of integrating over spins in a sphere\cite{genkin,summation}, wire, slab, or semi-infinite plane\cite{slab}. Discrepancies\cite{qiang} were found between their results and the experiments\cite{qiang, lopez, toro, du}. We believe the root of this inconsistency lies in their failure to acknowledge the mediating role of the carriers $\it {inside}$ nanoparticles. 

This neglect is particularly fatal when the medium is semiconducting because, while the previous approach would assume a ferromagnetic coupling for $\it {all}$ pairs of spins, the revised formula actually allows the possibility of being antiferromagnetic if only the pair distance varies by as little as a quarter of $5\sim 10$A - typical Fermi wavelength for metals. We shall focus on the realistic system of Mn-rich Mn(Ga)As nanoclusters in GaAs\cite{boeck,gaas} in this work, and argue that this oscillatory feature is intimately related to the spin-glass-like behavior often observed in this kind of system\cite{gaas,glass} besides the already been proposed temperature effect\cite{temp}. Armed by this knowledge, we explore the prospect of using the nanoclusters to achieve a higher Curie temperature.

\begin{figure}[h!]
\includegraphics[width=0.4\textwidth]{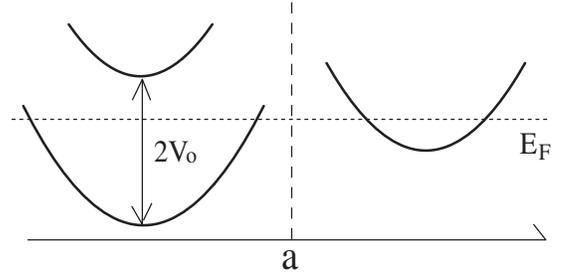}
\caption{\small Band structure inside the cluster is split by the Zeeman energy $2V_{0}$ due to the internal field, while that in the matrix is assumed to remain a single band.}
\label{fig:kf}
\end{figure}

When the long-range magnetic order is established among nanoclusters,
we have to consider the Zeeman splitting for the charge carriers.
Since the internal field in Mn(Ga)As is known to convert its band structure into half-metallic\cite{first}, we duly assign two bands for the carriers inside the cluster,
as shown schematically in Fig.\ref{fig:kf}, and different Fermi momenta $k_{F\sigma}$ can be defined for each spin $\sigma$. In contrast, the splitting in the medium is relatively weak and can be neglected\cite{mypp}.
Let us first calculate the indirect coupling between a spherical magnetic nanoparticle and a single impurity spin.
The general form of the electron eigenfunction is easiest expressed in spherical coordinates with the origin set at its center.
Symmetry allows us to separate the variables into
\begin{equation}
\psi_{lm\sigma}=S_{l\sigma}(r)Y_{lm}(\theta,\phi)
\label{eq:psi}
\end{equation}
where the radial part consists of the spherical Bessel functions, $j_{l}$ and $n_{l}$. Since the latter diverges at the origin,
we have to exclude it for $r\le a$:
\begin{equation}
S_{l\sigma}(r)=
\begin{cases}
j_{l}(k_{\sigma}r),  &\text { if $r\le a$};\\
A_{l\sigma}j_{l}(k^\prime r)+B_{l\sigma}n_{l}(k^\prime r), &\text{ if $r\ge a$}
\end{cases}
\end{equation}
where the three momenta are related by
\[\frac{k^{\prime 2}}{2m}=\frac{k_{\uparrow}^{2}}{2m}+V_{o}=\frac{k_{\downarrow}^{2}}{2m}-V_{o}\]
and the coefficients can be determined by matching the boundary condition as:
 \begin{align}
 \notag
&A_{l\sigma}=(k^\prime a)^{2}\left[\frac{k_{\sigma}}{k^\prime}j_{l+1}(k_{\sigma}a)n_{l}(k^\prime a)-j_{l}(k_{\sigma}a)n_{l+1}(k^\prime a)\right]\\
\notag
&B_{l\sigma}=(k^\prime a)^{2}\left[ j_{l}(k_{\sigma}a)j_{l+1}(k^\prime a)-\frac{k_{\sigma}}{k^\prime}j_{l+1}(k_{\sigma}a)j_{l}(k^\prime a)\right]
\end{align}

Finally, after properly normalizing the eigenfunction in Eq.(\ref{eq:psi}) by
$\sqrt{N_{l\sigma}}\equiv 1/\sqrt{A_{l\sigma}^2+B_{l\sigma}^2}$, we denote it by $\Psi_{k^\prime lm\rho}$.
Spatial variation of the magnetization can be calculated from the
eigenfunction obtained above:
\begin{equation}
\langle M(\overrightarrow{R})\rangle=\frac{s}{V}\sum_{k'<k'_{F}}\sum_{lm}
\big[\Psi^{*}_{k'lm\uparrow}\Psi_{k'lm\uparrow}-\Psi^{*}_{k'lm\downarrow}\Psi_{k'lm\downarrow}\big]
\label{eq:firstM}
\end{equation}
where $s$ is the spin of carriers and the expectation value is calculated in normal ordering.
One fact that comes in handy is that these nanoparticles shall be sparsely spaced in the matrix,
namely, their seperation \emph{R} is much greater than their size \emph{a}.
This allows us to take the asymptotic limit of the special function by Taylor expand and retain just
the lowest order term in $a/R$:
\begin{equation}
\langle M(R)\rangle\approx 2s k'^{3}_{F}\sqrt{Q^{2}+W^{2}}\frac{\cos(2k_{F}^\prime R+\phi(k'_{F}))}{(2k_{F}^\prime R)^{3}}
\label{eq:rho2}
\end{equation}
where $\phi=-\arctan(Q/W)$ and
$Q(k^\prime_{F})$ and $W(k^\prime_{F})$ are constants that depend on the parameters $(k^\prime_{F},a,V_{o})$:
\begin{align}
\notag
Q(k^\prime_{F})&=\Big[\frac{-A_{l\uparrow}^{*}A_{l\uparrow}+B_{l\uparrow}^{*}B_{l\uparrow}}{N_{l\uparrow}}-\frac{-A_{l\downarrow}^{*}A_{l\downarrow}+B_{l\downarrow}^{*}B_{l\downarrow}}{N_{l\downarrow}}\Big]\\
W(k^\prime_{F})&=\Big[\frac{A_{l\uparrow}^{*}B_{l\uparrow}+A_{l\uparrow}B_{l\uparrow}^{*}}{N_{l\uparrow}}-\frac{A_{l\downarrow}^{*}B_{l\downarrow}+A_{l\downarrow}B_{l\downarrow}^{*}}{N_{l\downarrow}}\Big]
\end{align}

Eq.(\ref{eq:rho2}) resembles the usual RKKY formula and is similar to the previous result\cite{genkin,skomski,summation,slab}
which integrated over all pairs of spins but did not include the contribution of the carriers inside the clusters. While it is not surprising to retain the oscillation with respect to $2k_{F}R$, we find the energy obtained from Eq.(\ref{eq:rho2}) only becomes identical to that of the previous work\cite{genkin,skomski,summation} when we turn off the Zeeman splitting inside the nanoparticles and set its dispersion relation to be the same as that in the medium. This implies that (1) our action of matching the boundary condition is equivalent to their carrying out the integration over spins inside the clusters; (2) the coefficient, $2\sqrt{Q^{2}+W^{2}}$, in Eq.(\ref{eq:rho2}) can be thought of
as an effective magnetic moment for the cluster; (3) whatever new physics that may come out of our inclusion of the inner carriers must lie in the phase term $\phi$.

By comparing Eq.(\ref{eq:rho2}) with the susceptibility expression, we can write down a Heisenberg-like Hamiltonian for clusters by reducing them effectively to point spins:
\begin{align}
H=\sum_{ij}\frac{2smk'^{4}_{F}J_{0}^{2}}{\pi^{3}\hbar^{2}}\frac{\cos(2k_{F}^\prime R+\phi_{i}+\phi_{j})}{(2k_{F}^\prime R)^{3}}\vec{S}_{eff,i} \cdot \vec{S}_{eff,j} 
\label{eq:rho3}
\end{align}
where $J_{0}S_{eff}={\pi^{3}\hbar^{2}}\sqrt{Q^{2}+W^{2}}/{m k'_{F}}$.
Detail derivations of Eq.(\ref{eq:rho3}) can be found in our previous article\cite{mypp}.

Finally, we are ready to study the realistic system of Mn-rich Mn(Ga)As nanoclusters\cite{gaas} in the GaAs matrix.
The fact\cite{gaas} that their interface is well-defined and both are zincblende without obvious dislocations lends credential to our treating the broundary as being step-like.
The carrier density and the effective mass in GaAs were roughly $10^{18}$ $cm^{-3}$ and 0.5$m_{e}$, and Mn density in the clusters was about 10\%\cite{gaas}.
When the Mn ions inside the nanoclusters are aligned, strength of the Zeeman splitting $V_{0}$ is estimated to be ``$N\beta S \times 10 \%$'' where $N\beta\sim 3.3$ ev and $S=5/2$ for Mn\cite{dietl,first}.

\begin{figure}[h!]
\includegraphics[width=0.4\textwidth]{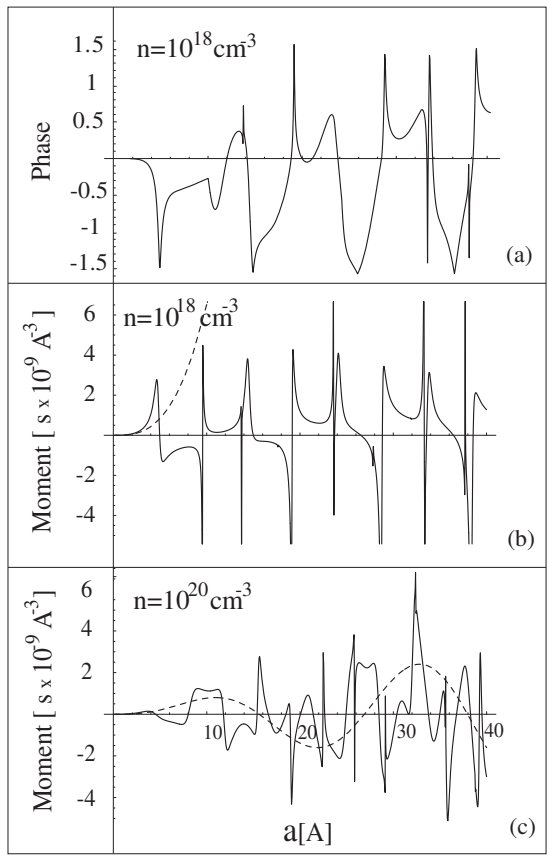}
\caption{\small (a): The phase term that appears in the inter-cluster coupling is plotted against the cluster radius.
Panels (b) and (c) show how the effective moment of each cluster varies with its size.
These properties are calculated at 20nm from the cluster center - the average spacing between neighboring clusters\cite{gaas}.
Dashed line is the moment obtained by the first-order Born approximation\cite{mypp} which is equivalent to the previous approach\cite{genkin,summation}, and the solid line is our result.}
\label{fig:radius}
\end{figure}

\begin{figure}[h!]
\includegraphics[width=0.4\textwidth]{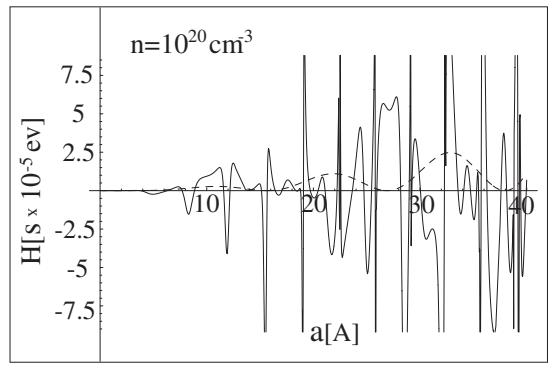}
\caption{\small Dashed line is the coupling energy obtained by the first-order Born approximation\cite{mypp} and the solid line is our result. The maximum of the latter can be enhanced to more than three times that by the previous approach\cite{genkin,summation}.}
\label{fig:cluster}
\end{figure}

The interesting phase $\phi$ and the effective moment are plotted against the cluster radius in Fig.\ref{fig:radius}.
The previous approach\cite{genkin,skomski} which predicted only ferromagnetism for this type of DMS irrespective of the cluster size, is inconsistent with the experiment\cite{gaas}.
In contrast, our result allows the coupling to oscillate with respect to the cluster size because the Fermi wavelength in the clusters is much shorter than in the matrix.
Resonance-like peaks arise in Fig.\ref{fig:radius}(b) and (c) whenever cluster radius satisfied the constraint of quantum interference.
Considering the fact that the size of these naturally formed Mn(Ga)As nanoclusters is bound to have a distribution, the spin orientation of each cluster is likely to be frustrated due to mixed preferences from different neighbors. We believe that this frustration may be responsible for the spin-glass-like behavior common to this type of system\cite{gaas,glass}.
In order to be applicable as DMS, the size distribution has to be narrowed to surpress the spin-glass-like phase. We propose that these nanoparticles be prepared separately and spatially arranged periodically on a substrate before being blanketed by layers of semiconductor. The coupling energy of this ideal array of nanoparticles is calculated and plotted as a function of the cluster size in Fig.\ref{fig:cluster}. It shows that the previous approach\cite{genkin,skomski} is likely to underestimate the optimal Curie temperature by more than three times. Since the embedded nanoclusters in many aspects share the same features as the magnetic dopants in DMS, it is expected that the system under investigation is also fit for a spin injector.

In conclusion, we calculated the indirect coupling between the sparsely-spaced  Mn-rich Mn(Ga)As nanoclusters in GaAs.
The conventional RKKY formula between point spins is modified by an extra phase term after we take into account the
differnt dispersion relation for carriers across the boundary of nanoclusters.
The spin-glass-like behavior observed experimentally can be explained by the mixture of ferromagnetic and antiferromagnetic coupling due to the nonuniformity of the cluster size, as opposed to the inhomogeneous spatial distribution of magnetic moments in metal\cite{inhomo}.
If the size distribution can be narrowed, DMS composed of magnetic nanoclusters have the potential of achieving a higher Curie temperature.

We benifit from discussions with Profs. H. H. Lin and D. W. Wang.
Support by the National Science Council in Taiwan under grant 95-2120-M007-008 is acknowledged.

%%%%%%%%%%%%%%%%%%%%%%%%%%%%%%%%%%%%%%%%%%%%%%%%%%%%%%%%%%%%%%%%%%%%%%%%%%%%%%%%
%%%%%%%%%%%%%%%%%%%%%%%%%%%%%%%%%%%%%%%%%%%%%%%%%%%%%%%%%%%%%%%%%%%%%%%%%%%%%%%%

%%%%%%%%%%%%%%%%%%%%%%%%%%%%%%%%%%%%%%%%%%%%%%%%%%%%%%%%%%%%%%%%%%%%%%%%%%%%%%%%
%%%%%%%%%%%%%%%%%%%%%%%%%%%%%%%%%%%%%%%%%%%%%%%%%%%%%%%%%%%%%%%%%%%%%%%%%%%%%%%%
%%%%%%%%%%%%%%%%%%%%%%%%%%%%%%%%%%%%%%%%%%%%%%%%%%%%%%%%%%%%%%%%%%%%%%%%%%%%%%

\end{document}